\documentclass[twocolumn]{aastex631}

\shorttitle{Mars as an Exoplanet}
\shortauthors{Stephen R. Kane et al.}

\begin{document}

\title{Mars as an Exoplanet: Lessons from a Planet at the Edge of Habitability}

\author[0000-0002-7084-0529]{Stephen R. Kane}
\affiliation{Department of Earth and Planetary Sciences, University of California, Riverside, CA 92521, USA}
\email{skane@ucr.edu}

\author[0000-0002-5644-7069]{Paul K. Byrne}
\affiliation{Department of Earth, Environmental, and Planetary Sciences, Washington University, St. Louis, MO 63130, USA}

\author[0000-0001-5592-6220]{Skylar D'Angiolillo}
\affiliation{Department of Earth and Planetary Sciences, University of California, Riverside, CA 92521, USA}

\author[0000-0002-0139-4756]{Michelle L. Hill}
\affiliation{Department of Earth and Planetary Sciences, University of California, Riverside, CA 92521, USA}
\affiliation{Department of Earth and Planetary Sciences, Stanford University, Stanford, CA 94305, USA}

\author[0009-0006-9233-1481]{Emma L. Miles}
\affiliation{Department of Earth and Planetary Sciences, University of California, Riverside, CA 92521, USA}

\author[0000-0001-8932-368X]{David A. Brain}
\affiliation{Laboratory for Atmospheric and Space Physics, University of Colorado, Boulder, CO 80304, USA}

\author[0000-0002-7463-9419]{Shannon M. Curry}
\affiliation{Laboratory for Atmospheric and Space Physics, University of Colorado, Boulder, CO 80304, USA}

\author[0000-0002-4834-2021]{Joana R.C. Voigt}
\affiliation{Department of Earth and Planetary Sciences, University of California, Riverside, CA 92521, USA}

%%%%%%%%%%%%%%%%%%%%%%%%%%%%%%%%%%%%%%%%%%%%%%%%%%%%%%%%%%%%%%%%%%%%

\begin{abstract}

Mars is the Solar System's canonical small, rocky planet that
transitioned from early geologic activity and surface liquid water to
a cold and arid planet with a thin, cold, CO$_2$-dominated
atmosphere. The evolution of Mars, in the context of such planetary
parameters as size, mass, atmosphere, insolation flux, magnetosphere,
and impact history, harbor important diagnostics regarding the
development and sustainability of habitable surface conditions. In
this work, we synthesize how the study of Mars contributes to our
understanding of exoplanet processes, such as volatile delivery and
loss, photochemistry, climate evolution (including CO$_2$ condensation
and atmospheric loss), obliquity forcing, planetary architecture, and
the role of intrinsic magnetism. We also evaluate optimal methods and
prospects for detecting and characterizing potential Mars analogs
beyond the Solar System. We focus on relevant results from planetary
missions (e.g., Mars Reconnaissance Orbiter, MAVEN, Mars Science
Laboratory, Mars2020) and observational studies of exoplanet
atmospheres with the James Webb Space telescope (JWST) and future
facilities. Through the convergence of these parallel pathways of
inquiry, we describe the primary science questions and suggested
avenues for characterizing small rocky planets that lie at the edge of
potentially habitable conditions.

\end{abstract}


%%%%%%%%%%%%%%%%%%%%%%%%%%%%%%%%%%%%%%%%%%%%%%%%%%%%%%%%%%%%%%%%%%%%

\section{Introduction}
\label{intro}

Exoplanet discoveries have advanced tremendously over the past few
decades, including the exploration of the demographics into the
terrestrial regime \citep{ford2014,winn2015,christiansen2025}, and
even those worlds that lie within their system's Habitable Zone (HZ)
\citep{kasting1993a,kane2012a,kopparapu2013a,kopparapu2014,kane2016c,hill2018,hill2023}. Results
from exoplanet surveys have demonstrated that small planets are more
common than their giant planetary cousins
\citep{howard2010b,dressing2015b}, creating an interesting comparison
to the Solar System's architecture
\citep{martin2015b,horner2020b,raymond2020a,kane2021d}. Yet, despite
their relatively high occurrence, the climates, volatile budgets, and
long-term habitability of small terrestrial planets remain poorly
understood \citep{tasker2017,adams2025a,apai2025}. Venus, Earth, Mars,
and even the Moon each underwent distinct volatile, tectonic, and
atmospheric trajectories despite sharing the same stellar environment,
illustrating that planet size alone does not uniquely determine
planetary evolution \citep{jakosky2025b,kane2021d}. A complete
framework for interpreting rocky exoplanets will ultimately require
integrating lessons from all of these bodies. With that broader
context in mind, we focus here on Mars because its sub-Earth mass,
thin atmosphere, and rich in-situ dataset make it particularly
instructive for the low-mass, volatile-poor regime that many detected
rocky exoplanets may inhabit. In that sense, Mars offers a uniquely
data-rich end-member: a sub-Earth-mass planet (0.107~$M_\oplus$,
0.53~$R_\oplus$) that lost most of its atmosphere and surface water
early, leaving isotopic fingerprints in the residual gases and surface
record \citep{hu2015c,jakosky2017a,jakosky2021}. Mars thus grounds our
models of escape physics, volatile cycling without plate tectonics,
and climate feedbacks that can drive atmospheres toward
collapse-phenomena anticipated on many close-in M-dwarf planets.

Mars occupies a singular position in comparative planetology: it is
simultaneously a well‐observed planetary neighbor with a rich rock and
climate record, and a physical regime (low gravity, thin atmosphere,
volatile sensitivity) that is highly relevant to the smallest rocky
exoplanets we are currently able to detect. Orbital spectroscopy and
in situ exploration reveal a basaltic crust variably altered by water,
with widespread phyllosilicates and sulfates that encode a transition
from early, more clement conditions to later aridity
\citep{ehlmann2014}. Stratigraphic and geochemical observations at
Gale crater demonstrate sustained lacustrine deposition with redox
gradients, pointing to environments that were habitable (and
chemically diverse) for geologically significant intervals
\citep{hurowitz2017}. At the same time, Mars demonstrates how small
planetary size and gravity shape climate evolution. Crustal
magnetization patterns measured by Mars Global Surveyor imply an
early, now‐extinct global dynamo \citep{acuna1999,connerney1999}, with
isotopic signatures such as $^{40}$Ar/$^{36}$Ar and D/H evidencing
extensive atmospheric escape thereafter. The Mars Atmosphere and
Volatile Evolution (MAVEN) mission quantified present‐day loss
channels (H, O, and ions) and, by integrating over enhanced early
solar activity, showed that escape plausibly removed a large fraction
of an early, thicker atmosphere \citep{jakosky2018b}. Together, the
geology and atmospheric evolution record of Mars offers a deeper
understanding of fundamental processes that are directly transferable
to interpreting surface-interior-atmosphere coupling on small
exoplanets \citep{shoji2014b,khuller2024,jakosky2025b}.

Furthermore, Mars' climate physics maps naturally onto exoplanet
modeling regimes of interest. Its CO$_2$–H$_2$O photochemistry and
HO$_x$–controlled oxidant budgets connect to generalized small‐planet
photochemistry \citep{lefeuvre2009c}; chaotic obliquity cycles and
dust–radiation feedbacks offer a template for forcings that may
episodically reshape sub-Earth climates
\citep{ward1973b,touma1993,jakosky1995a,forget2006,kane2025d}; and its
transition from wetter Noachian–Hesperian to cold–dry Amazonian
environments provides boundary conditions for testing long‐term
habitability on small worlds \citep{nair1994,zahnle2008d}. Mars is
also an important case-study in terrestrial atmospheric retention; an
aspect that has greatly shaped the early habitable history of the
planet and has great relevance to similarly sized exoplanets
\citep{jakosky1994b,chassefiere2004c,tian2009a,lammer2013a,dong2018b,basak2021}. The
atmospheric loss processes that have and continue to occur at Mars
thus inform the development of models that may be applied to a variety
of exoplanet contexts \citep{lammer2008,owen2019a}. These models are
particularly applicable to the TRAPPIST-1 system
\citep{roettenbacher2017,dong2018a}, whose relatively small size,
proximity to the host star, and age, may have resulted in their
atmospheric desiccation
\citep{greene2023,zieba2023,piauletghorayeb2025}, underscoring the
potential atmospheric vulnerability around active M-dwarfs. These
planetary properties, anchored by laboratory, orbital, and surface
measurements, make Mars an indispensable analog for connecting
process-level planetary science to the maturing exoplanet census of
Mars-sized and sub-Earth planets.

In this paper, we outline the properties of Mars in the context of an
exoplanet analog, and its importance in understanding the evolution of
planetary processes and surface habitability. Section~\ref{prop}
outlines the major features of Mars in comparison to Earth, and the
properties that have played a major role in its evolution and past
habitability. Section~\ref{demo} provides an analysis of the current
exoplanet demographics, and example discoveries that may fall into the
Mars analog category. Section~\ref{det} quantifies the detection
limits of exoplanet discovery methods for Mars analogs, including
diagnostics of current techniques and predictions for future
missions. Section~\ref{hab} focuses on the habitability prospects of
Mars analog exoplanets and their utility in testing the role of
various planetary processes that govern the sustainability of
temperate surface conditions. Section~\ref{discussion} discusses
various implications of Mars for interpreting exoplanet data, and
outlines observational discovery strategies optimized for Mars‐like
exoplanets, with special emphasis on systems hosting Mars‐size
candidates. We provide concluding remarks and suggestions for future
work in Section~\ref{conclusions}.

%%%%%%%%%%%%%%%%%%%%%%%%%%%%%%%%%%%%%%%%%%%%%%%%%%%%%%%%%%%%%%%%%%%%

\section{Mars: Fundamental Properties}
\label{prop}

\begin{figure*}
  \begin{center}
    \includegraphics[width=\linewidth]{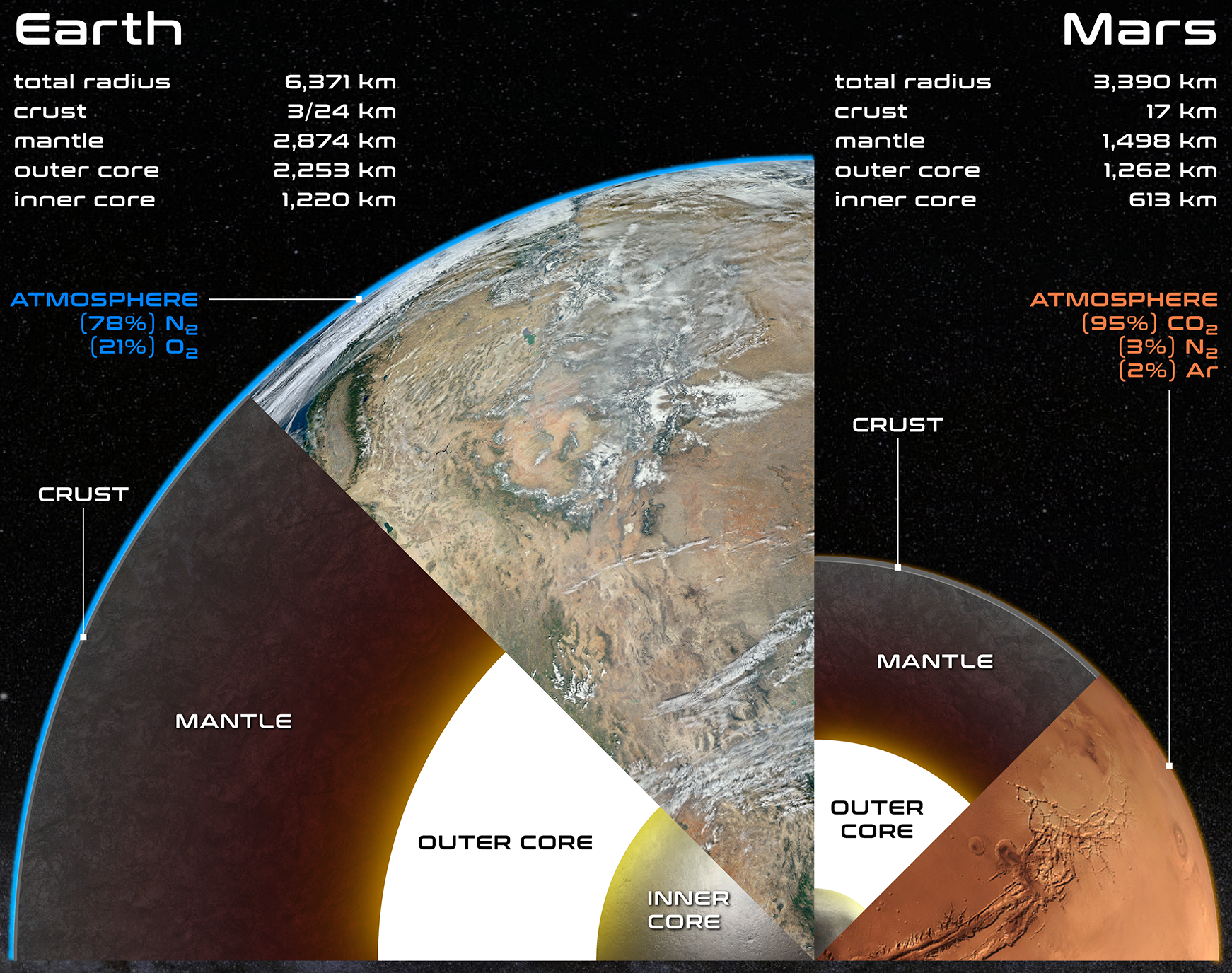}
  \end{center}
  \caption{Schematic cross sections of Earth and Mars, showing the
    major internal components and atmospheric components, to
    scale. For simplicity, oceanic and continental crust for Earth are
    not distinguished, nor is the interior structure of Earth's mantle
    shown.}
  \label{fig:prop}
\end{figure*}

Exoplanet studies often use Earth properties as standard units of
measurements, particularly for those relevant to describing the
capabilities of exoplanet detection methods. Mars has numerous
properties that are quite similar to Earth (such as obliquity and
rotation rate), and others that substantially diverge (such as mass
and surface pressure). Note that the present-day obliquity similarity
between Earth (23.44\degr) and Mars (25.19\degr) is a snapshot of a
chaotic dynamical history; as discussed in Section~\ref{geo}, Mars'
spin-axis tilt has ranged from $\sim$0\degr to $>$60\degr on
timescales of $\sim$$10^5$--$10^6$~yr, meaning the current value is
not representative of its long-term climate forcing. Shown in
Figure~\ref{fig:prop} and Table~\ref{tab:prop} are direct comparisons
of Earth and Mars properties, atmospheric composition, interior
structure, and various orbital and intrinsic parameters. Examining the
commonalities allows an assessment of scaling factors that can aid in
both the detection prospects and diagnosis of the potential Earth and
Mars exoplanet analogs.

\begin{deluxetable}{lrr}
\tablecaption{\label{tab:prop} Properties of Earth and Mars.}
\tablehead{
  \colhead{Property} & 
  \colhead{Earth} & 
  \colhead{Mars}
}
\startdata
Mass (\% Earth)             & 100        & 11 \\
Radius (\% Earth)           & 100        & 53 \\
Surface gravity (\% Earth)  & 100        & 38 \\
Insolation flux (\% Earth)  & 100        & 44 \\
Solar day (\% Earth)        & 100        & 103 \\
Surface pressure (\% Earth) & 100        & 0.63 \\
Semi-major axis (AU)        & 1.00       & 1.52 \\
Obliquity ($\degr$)         & 23.44      & 25.19 \\
Magnetic field (G)          & 0.25--0.66 & $\epsilon$ \\
Bond albedo                 & 0.316      & 0.25 \\
Geometric albedo            & 0.43       & 0.17 \\
Moons                       & 1          & 2 \\
Moon-planet mass fraction   & 0.01       & $\epsilon$
\enddata
\tablenotetext{}{$\epsilon$ indicates negligible quantity.}
\end{deluxetable}

%%%%%%%%%%%%%%%%%%%%%%%%%%%%%%%%%%%%%%%%%%%%%%%%%%%%%%%%%%%%%%%%%%%%

\subsection{Formation and Orbit}
\label{form}

Isotopic chronometers and dynamical models together suggest that Mars
formed rapidly and then stalled at sub–Earth mass because its feeding
zone was truncated. Hafnium–tungsten measurements indicate that Mars
accreted most of its mass within $\lesssim$2–-4~Myr after
calcium-aluminum–rich inclusions (CAIs), consistent with a stranded
planetary embryo rather than the product of late giant impacts
\citep{dauphas2011b}. Two end-member architectures can reproduce the
Earth/Mars mass contrast and aspects of the asteroid belt: (i) the
Grand Tack, in which Jupiter’s inward migration to $\sim$1.5~AU and
subsequent reversal depleted solids beyond $\sim$1~AU and starved the
Mars region \citep{walsh2011c}; and (ii) ``annulus'' initial
conditions that confine most terrestrial material to
$\sim$0.7--1.0~AU, naturally producing an over-massive Earth/Venus and
an under-massive Mars \citep{hansen2009c}. Subsequent simulations show
that both families of models can match the Earth/Mars mass ratio while
reproducing the belt's excitation and compositional mixing
\citep{raymond2009c,obrien2014a}.

Today Mars orbits with semi-major axis $a \simeq 1.5237$~AU,
eccentricity $e \simeq 0.0934$, and inclination $i \simeq
1.85^{\circ}$ (J2000); these values and their uncertainties are
provided by modern, refereed planetary ephemerides (e.g., Jet
Propulsion Laboratory (JPL) Planetary and Lunar Ephemerides DE440 and
DE441) and are consistent over multi-decade fits to spacecraft and
radar tracking \citep{park2021}. Mars' comparatively large $e$ and $i$
among the terrestrial planets reflect long-term secular forcing
dominated by the giant-planet modes, and its orbit undergoes
substantial quasi-periodic and chaotic variations in $e$ and $i$ over
$10^6$–$10^8$ yr \citep{laskar1993a,laskar2004b}. Mars, in turn,
exerts its own influence within Solar System dynamical interactions,
playing a role within Earth's Milankovitch cycles
\citep{kane2025d}. Ensemble integrations of the full Solar System show
that although Mercury can occasionally be driven to extreme
eccentricities, the probability of catastrophic inner-planet
instability over the next $\sim$5~Gyr is at the percent level; Mars
remains dynamically long-lived aside from its chaotic secular
excursions and strongly chaotic obliquity history
\citep{laskar2009c,zeebe2015a,mogavero2021}.

%%%%%%%%%%%%%%%%%%%%%%%%%%%%%%%%%%%%%%%%%%%%%%%%%%%%%%%%%%%%%%%%%%%%

\subsection{Geological and Climatic Evolution of Mars}
\label{geo}

Orbital spectroscopy and in-situ analyses show extensive Noachian-age
alteration by liquid water: phyllosilicates followed by Hesperian
sulfates, consistent with a progressively drying planet. This
stratigraphic transition from "phyllosian" to "theiikian" provides a
timeseries of water–rock interaction, pH, and redox conditions
\citep{ehlmann2014,bibring2006}. The NASA Curiosity rover established
that an ancient, long-lived lacustrine system was once present at Gale
Crater with neutral pH, low salinity, and essential
nutrients—demonstrably habitable for chemolithoautotrophs. The
Perseverance rover confirmed a delta–lake system within Jezero Crater,
including flood deposits recording hydrologic variability
\citep{grotzinger2014b,mangold2021b}.

Paleomagnetic and crustal remanence data constrain the history of
Mars' global dynamo, though that history is more complex than a simple
early cessation. Analyses of Mars Global Surveyor data placed the end
of the dynamo in the Noachian, at $\sim$4.0--4.1~Ga, based on the
absence of strong crustal magnetization over the large impact basins
Hellas, Argyre, and Isidis \citep{lillis2013c}. However, MAVEN
magnetic field measurements over the $\sim$3.7~Ga Lucus Planum lava
flows reveal confined near-surface magnetization consistent with a
dynamo that was active near the Noachian--Hesperian boundary,
substantially later than previously thought
\citep{mittelholz2020a}. This finding implies that the dynamo may have
postdated much of the geomorphic and mineralogic evidence for surface
liquid water described above, complicating simple narratives in which
dynamo cessation triggers atmospheric loss and climate
decline. Furthermore, it remains unclear whether a global magnetic
field provides a net protective effect against atmospheric
escape. While an intrinsic dipole deflects some solar wind ion pickup,
it can also facilitate loss through polar wind outflow and cusp ion
escape, such that the presence of a magnetosphere does not guarantee
reduced atmospheric erosion \citep{gunell2018a,brain2013}. These
nuances are important for interpreting exoplanet scenarios where the
presence or absence of an intrinsic magnetic field is often assumed to
be a binary indicator of atmospheric retention.

Mars undergoes chaotic obliquity variations that modulate insolation
distribution, volatile redistribution, and atmospheric pressure via
seasonal and long-term CO$_2$ condensation-sublimation at the poles
and regolith \citep{laskar1993b,laskar2004b}. Without a large
stabilizing moon comparable to Earth's, Mars' spin-axis tilt has
ranged from $\sim$0\degr to $>$60\degr on timescales of
$\sim$$10^5$--$10^6$~yr \citep{ward1973b,touma1993}, driving
order-of-magnitude swings in polar insolation and seasonal CO$_2$
cycling \citep{jakosky1995a,forget2006}. At high obliquity, polar
volatiles migrate toward the equator and atmospheric mass may
temporarily increase, whereas at low obliquity, CO$_2$ can collapse
onto the poles and reduce surface pressure below the triple-point
threshold for liquid water. These mechanisms provide a natural
laboratory for atmospheric stability thresholds and collapse pathways
that offer insights relevant to sub-Earth exoplanets, including
tidally locked worlds where qualitatively similar, though dynamically
distinct, cold trapping of volatiles can occur
\citep{wordsworth2015a}. Mars also exerts its own secular forcing on
the inner Solar System, contributing to Earth's Milankovitch cycles
\citep{kane2025a}, further illustrating how planetary architecture
shapes long-term climate evolution.

Mars' thin CO$_2$ atmosphere (with O, CO, O$_2$, N-bearing species,
and HOx) is shaped by photochemical cycles that regenerate CO$_2$ from
CO/O and regulate odd-hydrogen chemistry
\citep{nair1994,krasnopolsky2006c}. The rapid photodissociation of
CO$_2$ by solar UV radiation produces CO and O, but catalytic cycles
involving odd-hydrogen species (H, OH, HO$_2$) efficiently recombine
these products and maintain the observed $\sim$95\% CO$_2$
abundance. This balance is sensitive to the water vapor profile, dust
loading, and UV flux (parameters that vary with orbital state and
atmospheric dust events) making the Martian photochemical system a
real-world testbed for photochemical models of thin CO$_2$ atmospheres
on rocky exoplanets \citep{lefeuvre2009c}. In the exoplanet context,
Mars photochemistry constrains the production rates of abiotic O$_2$
and O$_3$, which are key false-positive biosignature gases expected on
irradiated, low-outgassing worlds. The Martian example also provides
empirical constraints on the thermodynamic conditions under which
CO$_2$ condensation occurs, which are relevant to the nightside cold
traps of tidally locked exoplanets, where the atmospheric circulation
regime and heat-transport efficiency introduce additional controls on
collapse thresholds \citep{wordsworth2015a}.

%%%%%%%%%%%%%%%%%%%%%%%%%%%%%%%%%%%%%%%%%%%%%%%%%%%%%%%%%%%%%%%%%%%%

\subsection{Atmospheric Loss: Mechanisms and Constraints}
\label{loss}

The present-day Martian atmosphere provides direct evidence that small
rocky planets are highly susceptible to long-term volatile
loss. Although the present atmosphere of Mars is relatively thin, the
early atmosphere was likely far more substantial, since that is
required to account for the hosting of surface liquid water
\citep{pollack1987a,jakosky2001a,wordsworth2016b,kite2019a,warren2019b}. For
example, \citet{joiret2025b} estimated a conservative minimal mass of
the primordial Martian atmosphere that implies a surface pressure of
at least 2.9 bar. This primordial atmosphere, present during and
immediately following planetary formation, was itself shaped by early
volatile processing including hydrodynamic escape during magma-ocean
stages, isotopic exchange between the atmosphere and solidifying
mantle, and impact-driven loss and delivery
\citep{pepin1991,kasting1993b,elkinstanton2008a,pahlevan2007}. The
subsequent atmosphere that hosted the surface liquid water recorded in
the geologic record (beginning $\sim$4.0--4.3~Ga) may have differed
substantially from this primordial state.

More broadly, the atmospheric evolution of any planet depends
critically on the initial volatile inventory incorporated during
formation, which in turn reflects the composition, location, and
timing of accreting materials \citep{jakosky2023b}. Given that the
initial incorporation of volatiles into planets remains poorly
understood even within our own Solar System (where we have extensive
observations of planets, asteroids, and meteorites) the range of
possible atmospheric outcomes for Mars-sized exoplanets with unknown
formation histories is correspondingly large. This uncertainty in
initial conditions must be borne in mind when extrapolating from Mars
to exoplanet populations.

Lyman-$\alpha$ observations and analyses of combined data from MAVEN
and the Emirates Mars Mission (EMM) reveal that hydrogen escape—often
diffusion-limited—can be seasonally or dust-storm enhanced; large dust
events correlate with spikes in upper-atmosphere H and increased
escape, demonstrably linking climate and loss
\citep{chaffin2014,heavens2018,chaffin2021}. Recent observations from
MAVEN have provided direct quantitative constraints on the dominant
non-thermal escape pathways operating at Mars today, helping to anchor
escape models for small exoplanets subjected to elevated XUV and
stellar winds. For example, measurements of ion escape rates indicate
that atmospheric loss is strongly modulated by solar wind conditions,
including dynamic pressure and interplanetary magnetic field
orientation, with significant enhancements observed during space
weather events such as coronal mass ejections
\citep{jakosky2015c,jakosky2018b}. MAVEN observations further
demonstrate that pickup ion escape, in combination with solar
wind-driven sputtering, represents an ongoing mechanism for
atmospheric erosion in the absence of a global magnetic dynamo
\citep{brain2016}. Additionally, photochemical escape of oxygen,
driven by dissociative recombination of O$_2^+$ in the upper
atmosphere, currently exceeds ion escape as an O loss channel and can
operate regardless of whether a global dynamo is present
\citep{lillis2017}.

Enrichment of heavy isotopes ($^{38}$Ar/$^{36}$Ar, D/H,
$^{15}$N/$^{14}$N) from in-situ measurements indicates substantial
atmospheric loss by mass-selective processes \cite{jakosky2017a}. Such
isotopic forensics calibrate retrieval expectations for
exo-atmospheres once precision spectroscopy becomes feasible
\citep{mahaffy2013,webster2013a}. Early Mars likely experienced
hydrodynamic H escape during magma-ocean/outgassing phases
\citep{pepin1991,tian2009a} and episodic atmospheric loss due to large
impacts \citep{melosh1989a}; both processes leave distinct isotopic
and volatile-inventory signatures relevant to young exoplanets. These
results suggest that episodic climate variability may contribute to
long-term volatile depletion and provide an empirical benchmark for
evaluating atmospheric retention on Mars analog exoplanets,
particularly those orbiting active stars with elevated stellar wind
fluxes \citep{brain2026}.

%%%%%%%%%%%%%%%%%%%%%%%%%%%%%%%%%%%%%%%%%%%%%%%%%%%%%%%%%%%%%%%%%%%%

\subsection{Past Habitability: Duration and Conditions}
\label{pasthab}

Liquid water is the essential solvent for life as we know it, and the
history of its abundance on Mars is central to assessing planetary
habitability. The past presence of water, and thus potential habitable
environments, is recorded in the Martian geologic record through
geomorphological features such as catastrophic outflow channels,
fluvial channel systems and valley networks, sedimentary fan deposits,
and features potentially consistent with a former ocean. These
geomorphological indicators are strongly supported by orbital spectral
observations that reveal a wide variety of hydrated minerals and
mineraloids, including Fe/Mg-smectite clays, chlorites, kaolinite,
mica, hydrated silica, zeolites, sulfates, serpentine, and
carbonates. Together, these mineral assemblages indicate widespread
and diverse water–rock interactions over Mars’ history
\citep[e.g.,][]{bibring2006,milikan2008,mustard2008,ehlmann2009,ehlmann2010,ehlmann2011,wray2011b},
with Fe/Mg-smectite clays being the most abundant, followed by
chlorites.

Longer-lived and widespread liquid water on Mars most likely occurred
during the pre-Noachian to Noachian periods. Shorter-lived episodes of
liquid water at the surface and in the subsurface, along with
associated water–rock interactions, likely persisted into the
Hesperian and Amazonian periods and may have formed protected or
transient habitable environments
\citep{cockell2014a,westall2015}. Specifically, sedimentologic facies
and geochemical characteristics at both ongoing rover missions sites
show that lakes at Gale crater and deltas at Jezero crater persisted
for at least centuries to millennia (likely longer), with redox
couples (Fe/S) and nutrients sufficient for microbial metabolisms. The
preservation potential of fine-grained mudstones and deltaic foresets
informs taphonomic criteria for future biosignature searches on
exoplanets (e.g., depositional ``traps'' inferred from global
photometry and spectra) \citep{grotzinger2014a,mangold2021a}.

In addition to lacustrine settings, hydrothermal environments
represent prime targets in the search for biosignatures, as they
provide sustained heat and liquid water and can concentrate biological
materials. Moreover, the highest potential for biosignature
preservation is found in geologic materials that experienced limited
diagenesis and metamorphism, minimizing overprinting of primary
signatures \citep[e.g.,][Opal-A]{sun2018,voigt2024,millan2025}, as
well as limited exposure to the Martian surface and atmosphere, where
oxidizing conditions and radiation would otherwise degrade or destroy
the initial signature. The taphonomic criteria established by Mars
exploration (particularly the preservation potential of fine-grained
mudstones, deltaic foresets, and hydrothermally altered materials)
provide a transferable framework for identifying high-priority
biosignature targets on rocky exoplanets, where analogous depositional
``traps'' might be inferred from disk-integrated photometry and
spectra.

%%%%%%%%%%%%%%%%%%%%%%%%%%%%%%%%%%%%%%%%%%%%%%%%%%%%%%%%%%%%%%%%%%%%

\section{Exoplanet Demographics}
\label{demo}

\begin{figure*}
  \begin{center}
    \includegraphics[width=0.9\linewidth]{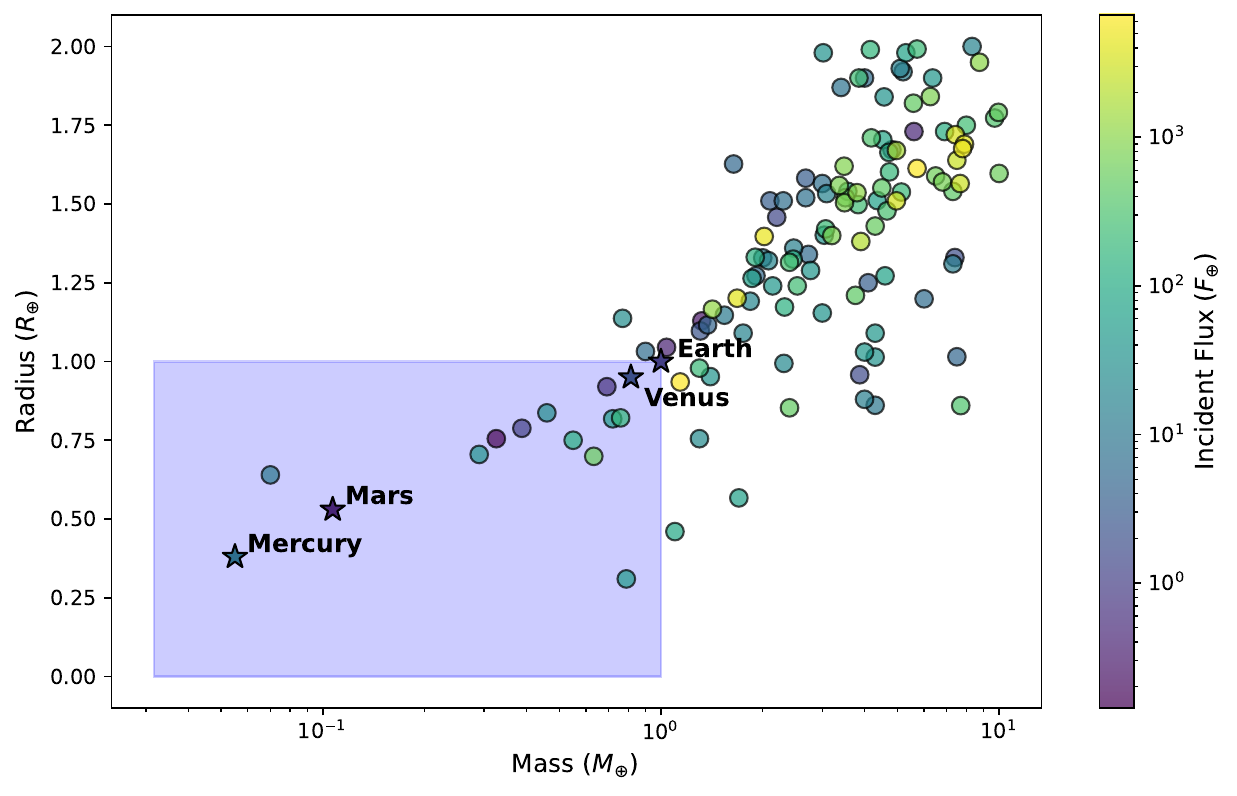}
  \end{center}
  \caption{Planetary mass and radius data for those confirmed
    exoplanets that have measurements extracted for both properties,
    extracted from the NASA Exoplanet Archive on 2025, December
    31. The data are color-coded in proportion to the flux received
    from their host stars. The Solar System terrestrial planets are
    shown as stars. The shaded region indicates the sub-Earth regime,
    defined as $M_p < 1$~$M_\oplus$ and $R_p < 1$~$R_\oplus$.}
  \label{fig:massrad}
\end{figure*}

As an exoplanet analog, Mars has numerous properties and processes
that are directly translatable to diagnosing planetary evolution in a
general planetary framework. For example, CO$_2$ frost cycling,
radiative–dynamical coupling in a tenuous atmosphere, and
dust–radiation feedbacks constrain climate stability models for
tidally locked terrestrial exoplanets, testing predictions of CO$_2$
atmospheric collapse thresholds \citep{wordsworth2015a}, although the
dynamical regimes differ and such extrapolations carry inherent
uncertainties associated with boundary conditions and processes that
may not be captured by models validated only under present-day Martian
conditions. Empirical atmospheric escape rates under weak gravity, and
their XUV/wind dependencies, partially calibrate models for small
exoplanets, including those around active M-dwarfs where non-thermal
processes can dominate, though the substantially harder XUV spectra
and more variable flaring environments of M-dwarfs shift the relative
weights of escape channels in ways that Mars alone cannot fully
constrain \citep{jakosky2018b,brain2026}. Noble-gas and hydrogen
isotope fractionation under known escape regimes calibrate how we read
future high-precision spectra of small exoplanet atmospheres
\citep{mahaffy2013}. Furthermore, Mars' photochemistry
(CO/CO$_2$/O$_2$/O$_3$/HOx) constrains oxidation states and potential
false positives (abiotic O$_2$/O$_3$) expected on irradiated,
low-outgassing exoplanets \citep{lefeuvre2009c}.

Empirical mass-radius measurements for small exoplanets provide the
critical mapping from the radius-rich transit census to physically
interpretable masses, bulk compositions, and volatile fractions
\citep{dorn2015,unterborn2023}. However, the terrestrial regime
remains strongly data-limited once one requires both the mass ($M_p$)
and radius ($R_p$) of the planet. Shown in Figure~\ref{fig:massrad}
are the mass and radii for all confirmed exoplanets with reported
masses of $M_p \le 10$~$M_\oplus$ and and radii of $R_p \le
2$~$R_\oplus$, as extracted from the NASA Exoplanet Archive on 2025,
December 31 \citep{christiansen2025}. The query includes all planets
with non-null mass and radius entries regardless of measurement
provenance (RV, TTV, or other), and no explicit precision cut is
applied; we note that the demographic shape in
Figure~\ref{fig:massrad} is sensitive to the inclusion of
lower-precision TTV-derived masses. This yielded a total of 120
exoplanets, shown in Figure~\ref{fig:massrad} as data points that are
color-coded by the calculated flux received by the planet ($F_p$). We
also include the Solar System terrestrial planets, shown as stars. The
shaded region indicates the area of mass-radius space that is
sub-Earth in value, and includes 11 exoplanets: TRAPPIST-1 h,
TRAPPIST-1 e, TRAPPIST-1 d, Kepler-138 b, L 98-59 b, K2-266 c,
Kepler-37 b, HD 23472 e, HD 23472 d, Kepler-20 e, and GJ 367 b. A
subset of these 11 exoplanets have been monitored using the
transmission spectroscopy capabilities of JWST or the Hubble Space
Telescope (HST), with no evidence found for the presence of
substantial atmospheres
\citep{damiano2022b,zhang2024c,glidden2025,piauletghorayeb2025}.

The relative scarcity of data within the shaded region is not
necessarily evidence that Mars analogs are intrinsically rare; rather
it is an imprint of survey selection effects. Transit surveys, such as
Kepler \citep{borucki2016}, preferentially populate the observed
$M_p$-$R_p$ plane at short periods and high $F_p$ because the
geometric transit probability scales roughly as $R_\star/a$, where $a$
is the orbital semi-major axis. Furthermore, detectability requires
sufficient signal-to-noise and multiple observed transits, driving
completeness sharply downward for sub-Earth radii at long periods
\citep{kane2008b,borucki2010a,mulders2018}. Occurrence-rate analyses
that correct for these effects demonstrate that small planets are
common in radius space \citep{fressin2013,dressing2015b}, yet these
radius-based demographics do not readily translate into a mass-defined
Mars frequency \citep{mulders2018}. Thus, the sub-Earth (shaded)
portion of Figure~\ref{fig:massrad} is disproportionately anchored by
highly irradiated transiting planets around low-mass stars and/or by
transit-timing variations (TTVs) in compact, near-resonant
architectures \citep{howard2010b,rogers2015a,winn2015}.

Even so, there are already several candidate systems in (or near) the
Mars‐size domain that serve as potential testbeds for Mars analogs in
radius, though their orbital and irradiation environments differ
substantially from Mars. The ultra‐compact Kepler‐42 (KOI‐961) system
hosts three sub‐Earths including a Mars‐sized planet ($R_p \approx
0.57$~$R_\oplus$) \citep{muirhead2012a}. Kepler‐444 contains five
sub‐Earths with radii spanning roughly Mercury to sub‐Venus sizes
\citep{campante2015}. Kepler‐37 includes a Moon‐to–sub‐Earth sequence
and a $\sim$0.74~$R_\oplus$ planet that lies between Mars and Earth in
size \citep{barclay2013a}. The mass of the Mars-sized exoplanet
Kepler-138 b was determined via TTVs, further demonstrating the
challenge in establishing the properties of sub-Earth size planets
\citep{jontofhutter2015}. More recently, the nearby L~98–59 system
features a transiting $\sim$0.84~$R_\oplus$ planet plus additional
terrestrial companions with precise masses and radii, enabling
comparative studies of interior–atmosphere diversity at sub‐Earth
scales \citep{demangeon2021a}. Furthermore, the validated Mars-size
planet candidate KOI-4777.01 presents an interesting case of how such
planets evolve in extreme flux environments around an M-dwarf
\citep{canas2022a}. These systems exemplify target classes where
Mars‐derived priors on volatile budgets, photochemistry, and escape
histories can be quantitatively exercised against current and
forthcoming observations.

%%%%%%%%%%%%%%%%%%%%%%%%%%%%%%%%%%%%%%%%%%%%%%%%%%%%%%%%%%%%%%%%%%%%

\section{Detecting Mars Exoplanet Analogs}
\label{det}

Here we discuss prospects for detecting Mars exoplanets analogs, and
the potential for characterization of their properties and
atmospheres.

%%%%%%%%%%%%%%%%%%%%%%%%%%%%%%%%%%%%%%%%%%%%%%%%%%%%%%%%%%%%%%%%%%%%

\subsection{Transits}

The transit technique of exoplanet discovery has thus far detected the
vast majority of known exoplanets \citep{christiansen2025}. The
photometric depth induced by a planetary transit scales as a simple
relationship between the radius of the planet and host star, as:
$\delta \approx (R_p/R_\star)^2$. For example, a Mars-size ($R_p
\approx 0.53$~$R_\oplus$) planet transiting a late M-dwarf host star
($R_\star \approx 0.12$~$R_\odot$) will yield a transit depth of
$\delta \sim 1.6\times10^{-3}$, or $\approx$1600~ppm. Although a
transit signature of this amplitude is readily accessible from both
ground and space-based facilities, such a planet-star configuration
represents the best-case scenario. A Mars-size planet transiting K
dwarf ($R_\star \approx 0.7$~$R_\odot$) or G dwarf ($R_\star \approx
1.0$~$R_\odot$) host star will yield transit depths of 49~ppm and
24~ppm, respectively. Such transit depths are challenging even from
space-based photometry, particularly in the face of intrinsic stellar
variability
\citep{aigrain2004a,ciardi2011,morris2020b,fetherolf2023b}. However,
as described in Section~\ref{demo}, several transiting Mars-size
planets have indeed been detected, albeit within the high incident
flux regime where mass measurements are enabled via radial velocities
(RVs) or TTVs within systems where orbital resonance is present
\citep{grimm2018,agol2021}.

The detection of true Mars analogs (at relatively low flux) will face
the challenge of building up signal-to-noise over multiple transits
over an extended period. Although results from the Transiting
Exoplanet Survey Satellite (TESS) have significantly added to the
known inventory of small planets around M-dwarfs, the 27-day sectors
observations decreases sensitivity to longer period planets
\citep{ricker2015,kane2021b,brady2022}. On the other hand, the
upcoming PLAnetary Transits and Oscillations of stars (PLATO) mission
is designed for long-baseline, high-precision photometry on bright
nearby stars \citep{rauer2014,rauer2025}. PLATO will continuously
monitor selected fields for 2–-3~years, enabling detection of
terrestrial planets with much longer orbital periods than those
represented in the TESS inventory, including Earth analogs around
Sun-like stars and potentially planets in the outer HZs of
M-dwarfs. By extending transit searches to periods of $\sim$1~year or
longer, PLATO will complement the results of the Kepler and TESS
missions, and may yield systems of multiple transiting terrestrial
planets in strong resonance configurations
\citep{heller2022b,matuszewski2023,boettner2024b}.

%%%%%%%%%%%%%%%%%%%%%%%%%%%%%%%%%%%%%%%%%%%%%%%%%%%%%%%%%%%%%%%%%%%%

\subsection{Radial Velocities}

Many of the initial exoplanet detections were the result of RV
observations through spectroscopic observations of bright stars in the
solar neighborhood
\citep{butler2006,schneider2011,christiansen2025}. The push towards RV
precisions that enable terrestrial exoplanet detection has required
the evolution of Extreme Precision RVs (EPRVs) and the diagnosis of
contamination from stellar activity
\citep{saar1997b,fischer2016,vanderburg2016b,cale2021}. Numerous EPRV
instruments have been developed and deployed, with most aiming for
precisions better than $\sim$50~cm/s over long periods of observation
\citep{gibson2016,jurgenson2016b,schwab2016,seifahrt2016b,pepe2021}.

\begin{figure}
  \includegraphics[width=\linewidth]{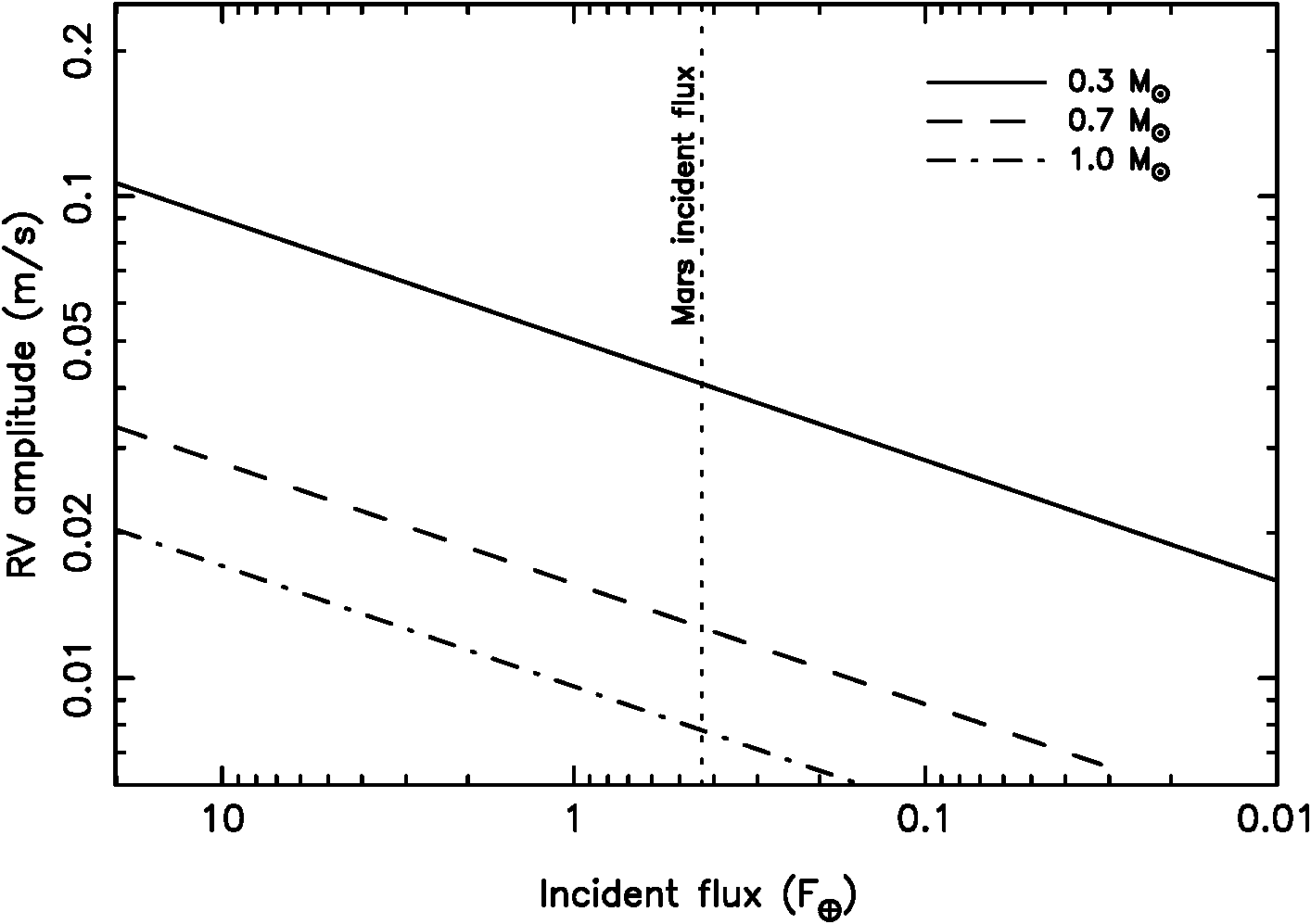}
  \caption{Predicted RV amplitude for a Mars-mass exoplanet as a
    function of incident flux, where the vertical dotted line
    indicated the incident flux for Mars ($\sim$0.43~$F_\oplus$). The
    calculations are shown for host stars of mass 0.3, 0.7, and
    1.0~$M_\oplus$.}
  \label{fig:rvplot}
\end{figure}

The RV amplitude of the Sun due to the gravitational influence of Mars
is $K \sim 7.8$~mm/s, which is substantially beyond the reach of
current EPRV instruments. By comparison, the RV amplitude induced on
the Sun by Earth is $K \sim 8.9$~cm/s. Figure~\ref{fig:rvplot} shows
the maximum (edge-on orbit) predicted RV amplitude for a Mars-mass
planet in a circular orbit as a function of incident flux for three
different host stars: an M-dwarf (solid line), a K-dwarf (dashed
line), and a G dwarf (dot-dashed line). The RV amplitude for a
Mars-mass planet at Mars equivalent flux around a K dwarf and M-dwarf
are $K \sim 1.3$~cm ($\sim$0.815~AU) and $K \sim 4.1$~cm/s
($\sim$0.185~AU), respectively. In order for a Mars-mass planet to
produce a similar RV amplitude to an Earth analog requires a flux of
10~$F_\oplus$ when orbiting an M-dwarf. Thus, the most promising
near-term opportunity for the RV detection of Mars-mass planets
requires targeting low-mass stars and short orbital periods.

%%%%%%%%%%%%%%%%%%%%%%%%%%%%%%%%%%%%%%%%%%%%%%%%%%%%%%%%%%%%%%%%%%%%

\subsection{Astrometry}

The astrometric method of exoplanet detection, relying on precise
measurements of stellar positions, has advanced considerably in recent
years, primarily through the data releases from the Gaia mission
\citep{perryman2014c,brandt2018,brandt2021a,brown2021}. For example,
the Gaia data have been used to refine the measured properties of
known exoplanets and their host stars
\citep{stassun2017,berger2018c,fulton2018b,delaverny2025} and have
demonstrated that some previously detected RV companions do not lie in
the planetary-mass regime \citep{keifer2019d}.

The semi-amplitude of the astrometric effect scales linearly with the
semi-major axis and mass of the planet, and thus favors massive
planets at large separations from the host star. For example, a
Neptune/solar analog located at a distance of 35~pc would produce an
astrometric semi-amplitude of 44~$\mu$arcsec \citep{kane2011d}, or
154~$\mu$arcsec if located at 10~pc. By comparison, a Mars/solar
analog located at 10~pc would induce a maximum astrometric
semi-amplitude of only 49~nas. A Mars orbiting at Mars-equivalent flux
around a K-dwarf and M-dwarf, also at 10~pc, would induce maximum
astrometric semi-amplitudes of $\sim$37~nas and $\sim$20~nas,
respectively. Thus, a Mars analog at 10~pc will produce an astrometric
signature that lies considerably below the noise floor of Gaia
($\gtrsim$10–20 $\mu$as) or any currently planned astrometric mission.

%%%%%%%%%%%%%%%%%%%%%%%%%%%%%%%%%%%%%%%%%%%%%%%%%%%%%%%%%%%%%%%%%%%%

\subsection{Gravitational Microlensing}

Gravitational microlensing probes exoplanets through the transient
magnification of a background ``source'' star as a foreground ``lens''
system closely aligns with the line of sight. In the planetary regime,
the observable signature is a short-lived perturbation to an otherwise
smooth point-lens light curve
\citep{paczynski1986d,mao1991a,gould1992g,paczynski1996a,gaudi2012}. Microlensing
is particularly sensitive to cold planets on wide orbits (typically
near the Einstein-ring scale) and to planetary systems at kpc
distances \citep{albrow2001c,cassan2012,gaudi2012}. This regime
naturally includes terrestrial and sub-terrestrial planets in
Mars-mass bins, provided that the planetary perturbation is
sufficiently well sampled and not erased by finite-source effects.

Modern ground-based microlensing planet searches rely on wide-field,
high-cadence monitoring of the crowded Galactic bulge from multiple
longitudes. Major survey facilities include the Optical Gravitational
Lensing Experiment (OGLE-IV) \citep{udalski2015b}, the Microlensing
Observations in Astrophysics survey (MOA-II) \citep{sako2008b}, and
the Korea Microlensing Telescope Network (KMTNet)
\citep{kim2016}. Despite these advances, detecting Mars analogs from
the ground remains difficult for three related reasons: (i) the
intrinsic rarity of microlensing alignments (necessitating very large
monitored star counts), (ii) the relatively short duration of low-mass
planetary perturbations, and (iii) systematic and sampling limitations
from atmospheric seeing, blending in dense fields, weather, and
diurnal gaps \citep{gaudi2012,zang2025}. In particular, small-planet
anomalies can be missed or poorly constrained when data are
interrupted during the perturbation \citep{zang2025}. Even so, the
ground-based low-mass frontier is approaching the Earth-mass regime in
favorable (typically high-magnification) events. For example,
\citet{zang2021c} reported KMT-2020-BLG-0414Lb with a planet-host mass
ratio $q \sim 10^{-5}$. By contrast, Mars analogs around typical bulge
lenses (0.3~$M_\odot$ at 6~kpc) correspond to $q \sim 10^{-6}$, where
the signal timescale shrinks and the detection cross-section drops by
an additional factor $\sim \sqrt{10}$. Consequently, present
ground-based statistical constraints on cold-planet demographics are
best developed at higher mass ratios rather than at Mars-like ratios.

The Nancy Grace Roman Space Telescope will conduct a dedicated
Galactic Bulge Time-Domain Survey (GBTDS) with a high-cadence,
near-infrared wide-bandpass sequence designed explicitly for
microlensing demographics, including sensitivity to low-mass planets
\citep{sajadian2021a,fatheddin2023b}. Space-based observations address
the principal ground limitations by providing weather-free, high duty
cycle during observing seasons, stable point-spread functions and
calibrated photometry, and substantially reduced crowding/blending
relative to typical ground seeing \citep{gaudi2012,terry2026}. Roman
yield forecasts discuss total event samples of order $\sim 5\times
10^4$ microlensing events over the nominal six-season survey,
depending on the event definition and impact-parameter cut
\citep{saggese2026}. Forecasts further indicate that Roman is expected
to detect bound planets above $\sim$~0.1~$M_\oplus$ (i.e., Mars mass)
in statistically useful numbers and maintain sensitivity down to
$\sim$~0.02$M_\oplus$ under favorable circumstances
\citep{penny2019}. Overall, gravitational microlensing remains one of
the most promising techniques for discovering cold, low-mass
exoplanets at wide separations, with the caveat of relatively brief
planetary deviations requiring sufficient observational cadence to
characterize the planetary system.

%%%%%%%%%%%%%%%%%%%%%%%%%%%%%%%%%%%%%%%%%%%%%%%%%%%%%%%%%%%%%%%%%%%%

\subsection{Direct Imaging and Thermal Emission}

Current ground-based direct-imaging surveys remain fundamentally
optimized for young, self-luminous giant planets rather than cold Mars
analogs. Even with extreme Adaptive Optics (AO), coronagraphy, and
modern post-processing, survey-grade near-IR performance is typically
in the $\sim$$10^{-6}$–-$10^{-7}$ contrast regime at arcsec
separations \citep{chomez2025}. Demographic constraints from large
ground surveys therefore concentrate on companions of a few to
$\gtrsim 10$~$M_J$ at multi-AU to $\sim$100~AU scales, not on
sub-Earth rocky planets \citep{nielsen2019c}. For calculating the
planet-star flux ratio of Mars analogs in reflected light, we adopt
the formalism of \citet{kane2010b,kane2011a}. Assuming a geometric
albedo of $A_g = 0.15$, a planetary radius of $R_p = 0.53$~$R_\oplus$,
a semi-major axis of $a = 1.52$~AU, and a Lambertian reflectance, we
calculate a planet-star flux ratio of $\epsilon = 1.1\times10^{-11}$,
which is several orders of magnitude fainter than present contrast
floors. Thermal emission has a less demanding intrinsic contrast
\citep{marley1999a,sudarsky2000,kane2011g,guzewich2020a}, which, for a
temperate Mars analog around a solar twin, is of order $\sim10^{-8}$
near wavelengths of 10~$\mu$m, but ground-based sensitivity at
thermal-IR wavelengths is critically limited by high thermal
backgrounds \citep{rousseau2024}.

Future space-based, high-contrast observatories are expected to shift
toward the feasibility of Mars-analog detectability with the Habitable
Worlds Observatory (HWO), explicitly conceived as a large
UV/optical/IR facility designed to directly image and
spectroscopically characterize potentially habitable planets
\citep{harada2024b,kane2024d,stark2024b,tuchow2024}. Post-processing
frameworks and operations concepts for a HWO-class coronagraph are
being developed around the need to measure flux ratios of
$\sim10^{-11}$ \citep{mcelwain2026}. For a Sun-like star at $d =
10$~pc, a Mars-orbit analog has an angular separation of $\theta =
0.152\arcsec$, which is outside the anticipated visible-light Inner
Working Angle (IWA). However, a robust detection will require a
substantially deeper search than for an Earth analog, as the reflected
light flux ratio of a Mars analog ($\epsilon \sim 10^{-11}$) is
roughly an order of magnitude fainter than for an Earth analog
($\epsilon \sim 10^{-10}$), requiring correspondingly longer
integration times. This implies that Mars-analog yields will likely be
driven by targeted deep observations of the nearest and most favorable
stars rather than uniform completeness across the full exoEarth sample
\citep{stark2024a}. Stellar-type and distance trade-offs may be
considered, meaning late-type stars offer higher reflected-light
contrast but smaller angular separations. These couplings motivate
yield optimization (and wavelength/IWA trade studies) as the
operational path to recovering Mars-analog completeness once HWO-class
stability and contrast are achieved \citep{stark2024a}. A dedicated
mid-IR space interferometer, such as the Large Interferometer For
Exoplanets (LIFE), would greatly expand the search domain, enabling
detection/characterization of hundreds of small planets, including
potential Mars analogs \citep{quanz2022a}.

%%%%%%%%%%%%%%%%%%%%%%%%%%%%%%%%%%%%%%%%%%%%%%%%%%%%%%%%%%%%%%%%%%%%

\subsection{Atmospheric Characterization}

Atmospheric characterization of Mars-like exoplanets is scientifically
compelling because even a tenuous CO$_2$-rich atmosphere can
profoundly affect surface habitability and attest to a planet's
volatile history \citep{tinetti2005,wolfe2024}. Alternatively, the
lack of a substantial atmosphere may be consistent with strong escape
mechanisms dominating over the atmospheric evolution, as has been
inferred from JWST observations in several cases
\citep{greene2023,zieba2023}. For example, a positive detection of
H$_2$O, O$_3$, or other photochemical tracers may indicate remnant
oceans or past high-energy irradiation, providing key benchmarks for
comparative planetology
\citep{schwieterman2018,fujii2018,ostberg2023c}. To investigate the
observable signatures, we simulated transmission and reflectance
spectra of a Mars analog using the Planetary Spectrum Generator
\citep[PSG;][]{villanueva2018b}. Each model uses the same Martian
planetary configuration. The model utilizes a surface pressure and
temperature of $\sim$6~mbar and $\sim$216~K, respectively. The bulk
composition of the model atmosphere is dominated by CO$_2$ (96.934\%),
with N$_2$ (1.897\%) and O$_2$ (0.144\%) as secondary constituents,
and trace amounts of CO, H$_2$O and O$_3$. The model atmosphere also
includes two aerosol species; dust (2.921~ppm) and water ice
(0.947~ppm), both of which have particle radii of 1.222~$\mu$m
\citep{millour2015,millour2024}. These aerosols, along with the
surface albedo ($A = 0.2448$) and emissivity ($\epsilon$ = 0.7555) are
wavelength-dependent, and the spectral signature's detectability is
influenced by the spectral energy distribution of the host star
\citep{shields2016b,wunderlich2019}. For solar-type stars, reflected
light dominates at shorter wavelengths, meaning that there are
stronger signals occurring at shorter wavelengths. On the other hand,
M-type stars dominate in the near-IR, and will therefore have stronger
features in that wavelength regime. This also means that surface and
atmospheric features in the near-IR will become more prominent
compared to that of a solar-type star. We note that these parameters
are specific to present-day Mars, and that variations in surface
pressure, albedo, or atmospheric composition (as might arise under
different stellar spectra or earlier evolutionary states) would affect
the depth and detectability of spectral features. For example,
increasing the dust opacities could raise the effective albedo and
reduce emission feature depths, while increasing the pressure would
broaden CO$_2$ absorption features and enhance transmission signals.

\begin{figure*}
  \begin{center}
    \includegraphics[width=\linewidth]{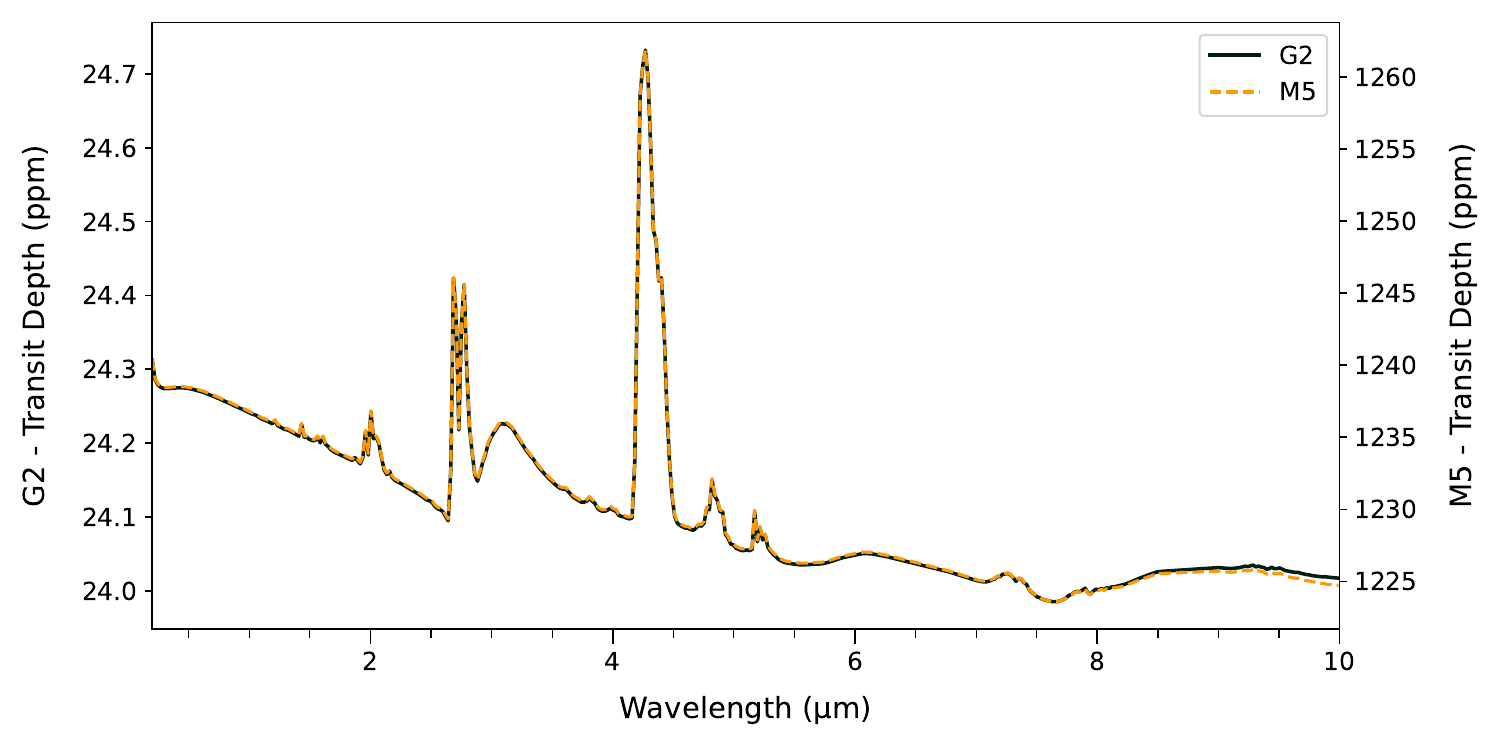}
  \end{center}
  \caption{Transmission spectrum of a Mars analog in an edge-on orbit
    around a host star located at a distance of 10~pc. The left
    and right vertical axes show the strength of the absorption
    features for a G2 dwarf and M5 dwarf host star, respectively.}
  \label{fig:transpec}
\end{figure*}

Figure~\ref{fig:transpec} shows the simulated transmission spectrum of
a Mars analog around a Sun-like star at a distance of 10~pc, over a
wavelength range of 0.2--10~$\mu$m (modeled on JWST/NIRSpec
PRISM). The left and right vertical axes shows the depth of the
absorption features for a G2 dwarf host and an M5 dwarf host,
respectively. Strong CO$_2$ absorption features appear at 2.7 and
4.3~$\mu$m, in analogy to those seen in Earth's spectrum, and span
several atmospheric scale heights. The corresponding transit depth
variation is extremely small: $< 10^{-6}$ for a Mars-size planet and
Sun-like star, making detection challenging even with JWST. Indeed,
\citet{kreidberg2025} noted that rocky exoplanet spectra require
multi-scale-height features for firm
detections. Figure~\ref{fig:transpec} demonstrates that, aside from
the strongest bands, most absorption features remain near or below the
noise floor, reflecting the high mean molecular weight and compact
atmosphere of Mars analogs \citep{lustigyaeger2023b,rigby2023a}.

\begin{figure*}
  \begin{center}
    \includegraphics[width=\linewidth]{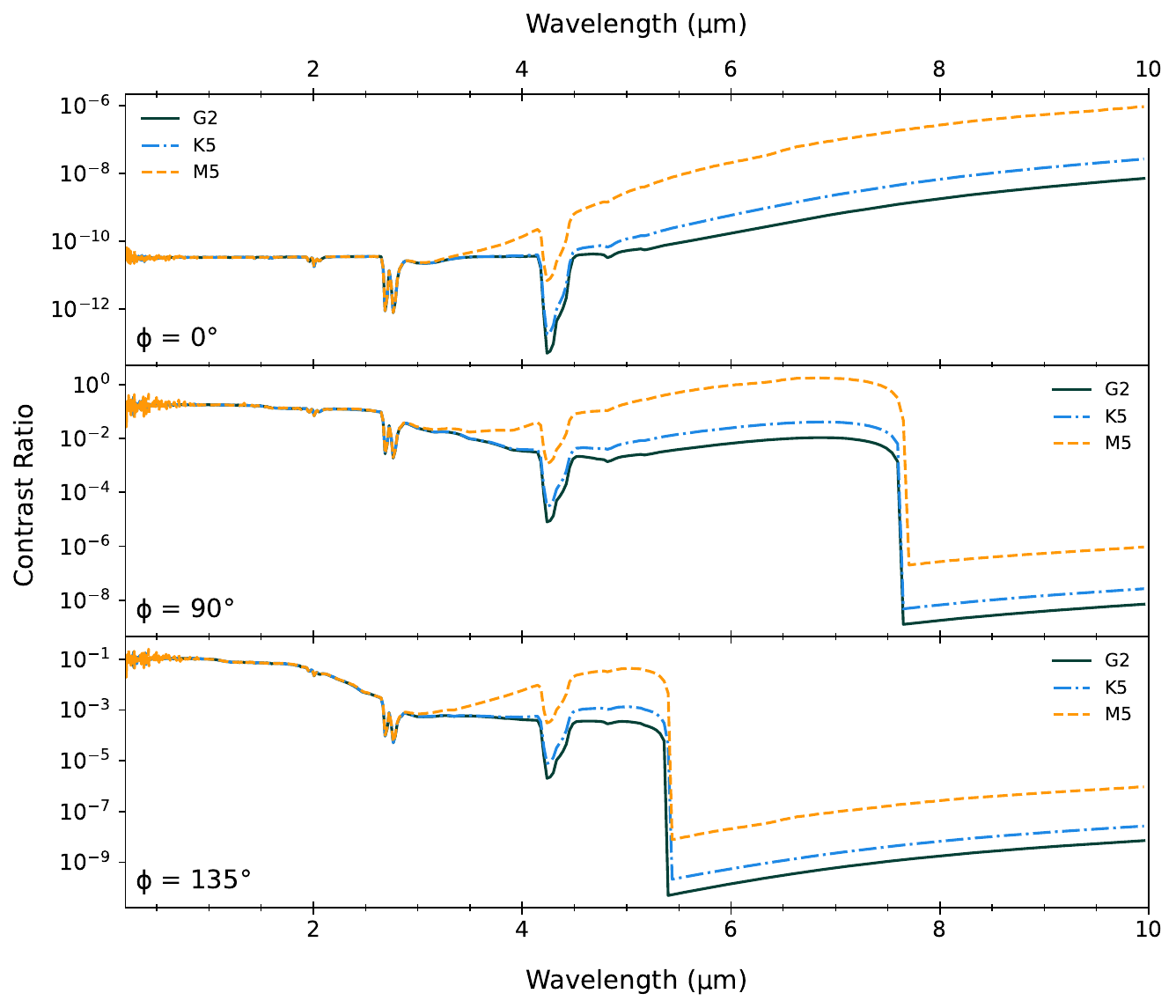}
  \end{center}
  \caption{Reflectance spectra for a Mars analog in an edge-on orbit around G2 dwarf, K5 dwarf, and M5 host stars located at a distance of 10~pcs. The top, middle, and bottom panels correspond to planetary phase angles of 0\degr, 90\degr, and 135\degr, respectively.}
  \label{fig:refspec}
\end{figure*}

Direct imaging in reflected light offers complementary spectral information. Figure~\ref{fig:refspec} displays model reflected spectra of a Mars analog observed at phase angles of 0\degr, 90\degr, and 135\degr, at a distance of 10~pcs, and assuming a 15~m coronagraph-equipped telescope over 0.2--10~$\mu$m \citep{kawashima2019a}, based approximately on the Large Ultraviolet Optical Infrared Surveyor (LUVOIR) A-VIS instrument \citep{reportluvoir}. The reflected spectra in Figure~\ref{fig:refspec} show modest signatures of CO$_2$ and surface mineral bands, with overall brightness declining at larger phase angles. As noted by \citet{roccetti2025b}, reflected-light spectra as a function of wavelength and phase angle encode atmospheric and surface properties (e.g., albedo variations and molecular absorptions such as O$_2$, H$_2$O, CH$_4$, CO$_2$). By comparing the three spectra in Figure~\ref{fig:refspec}, the full-phase (0\degr) spectrum is approximately uniformly bright across wavelength, while at 90\degr--135\degr the absorption features are weaker and the flux falls sharply at NIR wavelengths due to the reduction in Rayleigh and forward Mie scattering \citep{roccetti2025b,roccetti2025c}. These variations can be exploited to retrieve composition: for example, deep CO$_2$ bands at 4.3~$\mu$m and weaker H$_2$O or O$_3$ features will modulate the reflected color differently from bare surface \citep{garciamunoz2018a}. Thus, direct imaging at HWO-class contrasts ($\sim$$10^{-10}$) is expected to yield more abundant molecular information for nearby Mars analogs than transit spectroscopy alone.

%%%%%%%%%%%%%%%%%%%%%%%%%%%%%%%%%%%%%%%%%%%%%%%%%%%%%%%%%%%%%%%%%%%%

\section{Habitability of Martian Analogs}
\label{hab}

The long-term habitability of Mars-size exoplanets is controlled by
how quickly small rocky bodies cool, degas, and then lose their
atmospheres \citep{unterborn2022,hill2026}. As planetary radius
decreases, the smaller volume to surface area ratio allows conductive
cooling to occur more quickly, as too does a reduction in radiogenic
materials correlated to the mantle mass reduction of the planet. This
cooling has the effect of reducing convective vigor and decreasing
melt production, both of which may contribute to whether the planet is
able to start and/or maintain plate tectonics
\citep{oneill2007d,grott2011a,unterborn2022}. Analysis of Mars mantle
volatile abundances and atmospheric inventories supports these
theoretical predictions, demonstrating that initial volatile budgets
and subsequent outgassing histories are critical controls on long-term
atmospheric mass \citep{jakosky2023b}. In such bodies, most volatile
release occurs early, during magma–ocean solidification and the first
few hundred Myr of mantle overturn, followed by an order-of-magnitude
(or more) reduction in outgassing rates after $\sim$0.5–1.5~Gyr as
reducing mantle temperatures prohibit melt production
\citep{noack2017a,tosi2017}. Whether degassing can persist depends on
the efficiency of melt focusing through a thick, dehydrated
lithosphere and on the availability of carbonate reservoirs, and
models generally predict shorter-lived volcanic-degassing epochs for
Mars-mass planets than for Earth analogs
\citep{foley2018a,grott2011a}. These model predictions are consistent
with the volcanic history of Mars compared with, for example, that of
Earth \citep{byrne2020}. These interior trends also couple to magnetic
evolution: small cores cool rapidly, so dynamos are typically brief,
potentially further exposing atmospheres to stellar wind erosion once
magnetism wanes \citep{driscoll2013,lillis2013c,gunell2018a}.

Atmospheric retention on Mars-size worlds is therefore a race between supply (early and episodic volcanism, impacts) and loss (thermal and non-thermal escape). High XUV irradiation drives hydrodynamic escape of H (and in extreme cases, drag-off of heavier species), with escape efficiencies that peak during the first $10^8$–-$10^9$~yr when stars are most active \citep{tian2009b,murrayclay2009}. Non-thermal processes (ion pickup, sputtering, and photochemical escape) become dominant once the upper atmosphere is H-poor, especially without a persistent magnetosphere \citep{jakosky2018b}. Around active M-dwarfs, the extended pre–main–sequence luminosity plateau and strong space weather (flares, CMEs, winds) can desiccate and erode secondary atmospheres. For example, simple energy-limited estimates imply loss of multiple terrestrial ocean equivalents in $\lesssim 100$~Myr for close-in orbits unless outgassing is exceptionally vigorous or shielding is strong \citep{luger2015b,airapetian2017,dong2018a,bolmont2017a}. In addition, thin CO$_2$ atmospheres on tidally locked planets can collapse into cold traps if heat transport and background pressure are insufficient, setting a CO$_2$-stability floor that is particularly constraining for sub-Earth masses \citep{joshi1997d,heng2012c,wordsworth2015a}.

\begin{figure*}
  \begin{center}
    \includegraphics[width=0.9\linewidth]{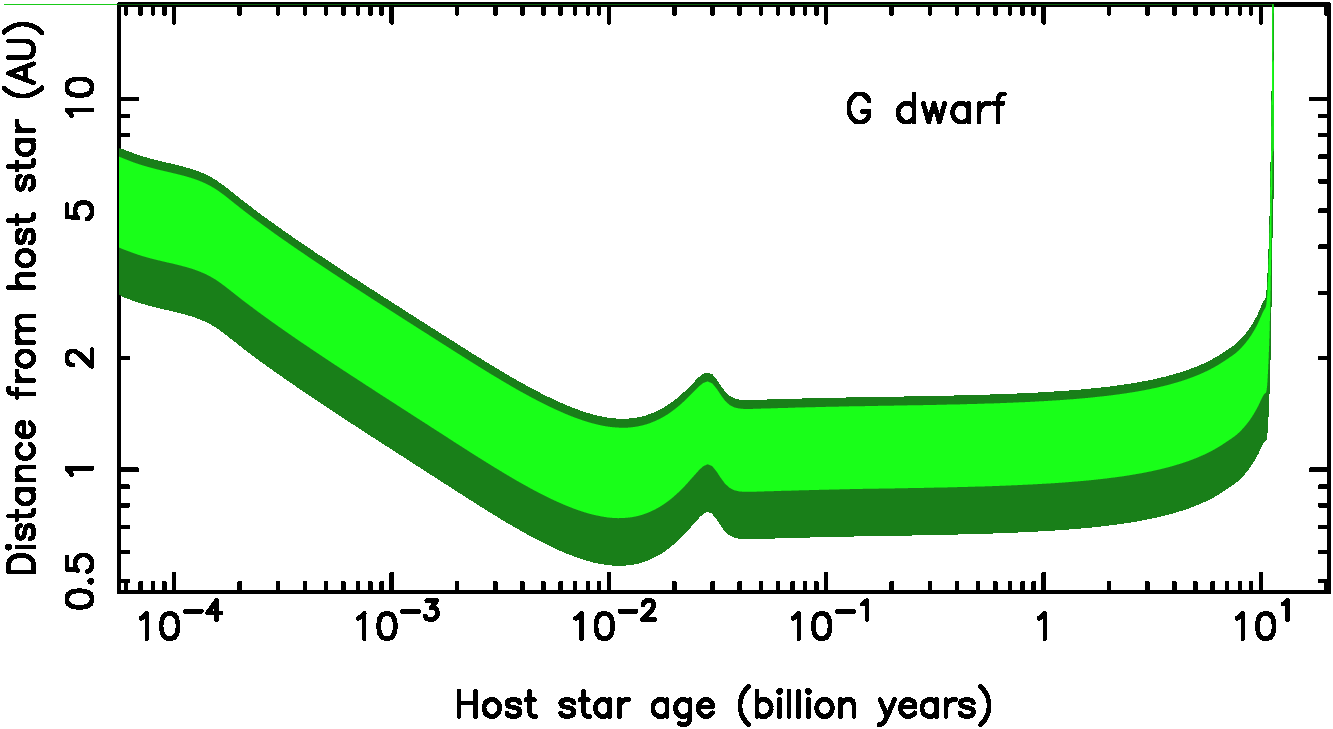} \\
    \includegraphics[width=0.9\linewidth]{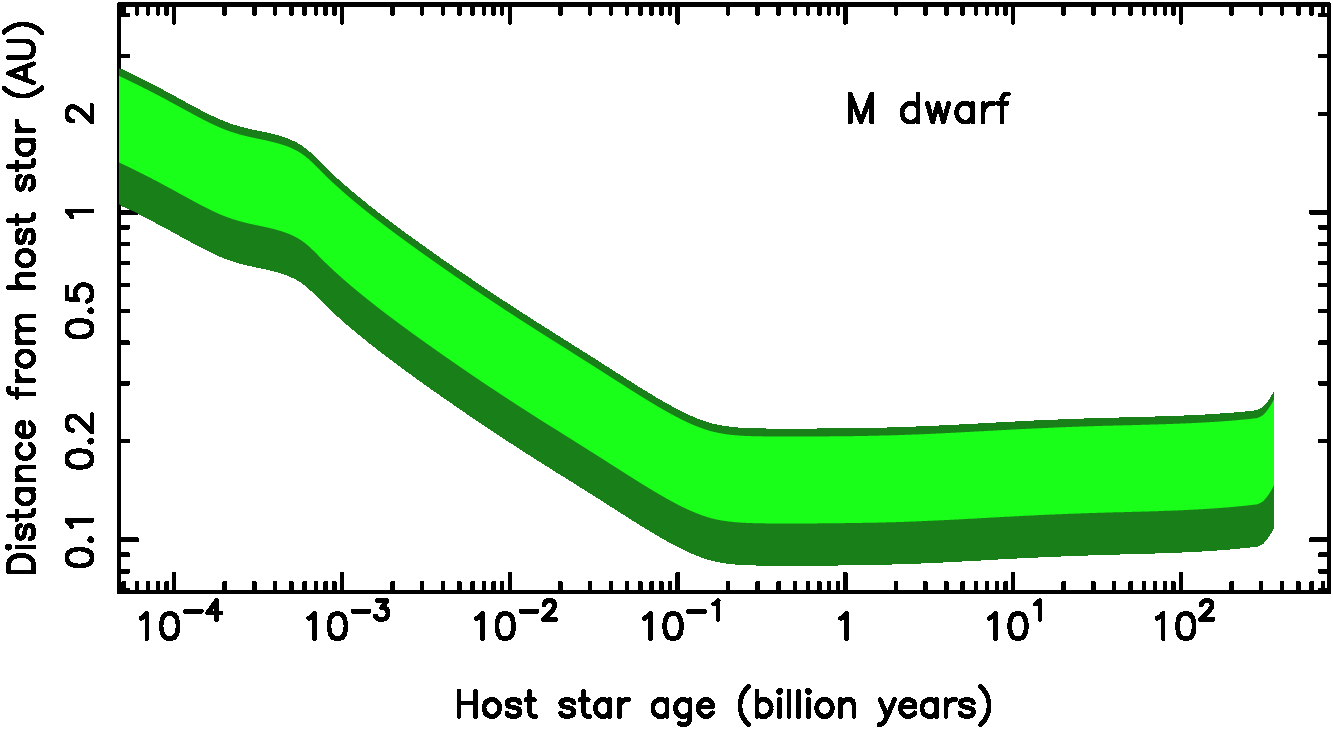}
  \end{center}
  \caption{Evolution of the HZ before and during the main sequence for an G dwarf (top) and M-dwarf (bottom). The extent of the HZ is shown in green, where light green is the conservative HZ and dark green is the optimistic extension to the HZ.}
  \label{fig:hzevolve}
\end{figure*}

Incident flux, spectral energy distribution, and orbital forcing all contribute to a planet's ability to host surface liquid water. Particularly for small planets, secularly maintained eccentricity in compact systems can raise the orbit–averaged instellation and produce sustained tidal heating that either prolongs volcanism and degassing or forces runaway volatile loss and interior desiccation \citep{jackson2008a,kane2012e,barnes2013a,vanlaerhoven2014,way2017a,kane2021a}. System architecture further modulates habitability by setting impact fluxes and volatile delivery (e.g., the presence and spacing of giant planets), by locking planets into resonant chains that maintain eccentricity and tides, and by altering long-term obliquity stability \citep{raymond2017b,turbet2018,kane2024a,kane2025a}. Regarding the incident flux, updated radiative–convective models place the classical main–sequence HZ for Sun-like stars at incident flux $F_p/F_\oplus \sim$~0.35--1.1, but the effective outer edge for Mars-size planets is closer to the star because (i) CO$_2$ condenses more readily, and (ii) heat transport is less effective in thin atmospheres \citep{kopparapu2013a,kopparapu2014,wordsworth2015a}. Conversely, transient greenhouse episodes (e.g., volcanic CO$_2$+H$_2$) can widen the viable window if outgassing sustains $\gtrsim$bar-level inventories \citep{ramirez2017b}. Host-star spectral type also matters through the XUV/particle environment that sculpts escape and through the near-IR weighting of the stellar spectrum, which enhances H$_2$O and CO$_2$ absorption for late-type stars, modifying HZ limits and the threshold for atmospheric collapse \citep{kopparapu2014,turbet2016}.

Shown in Figure~\ref{fig:hzevolve} are examples of the HZ evolution for a G dwarf (top panel) and M-dwarf (bottom panel). The stellar properties and evolutionary tracks were calculated using the MESA Isochrones \& Stellar Tracks (MIST) \cite{paxton2011,paxton2013,paxton2015,choi2016,dotter2016,paxton2018,paxton2019}. A solar metallicity was assumed for both stars, and stellar masses of 0.3~$M_\odot$ and 1.0~$M_\odot$ were used for the M-dwarf and G dwarf, respectively. The conservative HZ (CHZ) is bounded by calculated limits of runaway greenhouse and maximum greenhouse transitions, and the optimistic HZ (OHZ) boundaries are determined via empirical evidence of past surface liquid water on the surfaces of Venus and Mars \citep{kopparapu2013a,kane2016c}. The CHZ is shown as light green in Figure~\ref{fig:hzevolve}, and the OHZ extension to the HZ is shown as dark green. To calculate the HZ boundaries, we adopted the 0.1~$M_\oplus$ coefficients provided by \citep{kopparapu2014}, whereby the inner edge of the CHZ moves outward due to an expected increased atmospheric H$_2$O column depth. The timescales and distances shown in Figure~\ref{fig:hzevolve} demonstrate the extent to which the HZ evolution changes between G and M stars on the main sequence, particular the relative length of the pre-main sequence phase, which is considerably longer for M stars. Although the width of the HZ for M stars is quite narrow ($\sim$0.1--0.2~AU), it remains stable for many billions of years. However, the risk remains that the proximity of the Mars-mass planet to an extended period of increased stellar activity may result in a complete loss of atmosphere (see Section~\ref{loss}).

In summary, Mars-mass exoplanets are most likely to enjoy clement surface conditions early, when outgassing peaks, and later only if supplied by long-lived internal heat sources, tidal dissipation, or episodic volatile delivery. Long-term habitability is favored around quieter K/G stars at moderate incident flux, low-to-moderate eccentricity, and in architectures that neither overstimulate tides nor starve the planet of volatiles. By contrast, close-in M-dwarf planets require unusually robust volatile inventories, vigorous and persistent degassing, and/or magnetic and atmospheric protection to survive the pre-main-sequence and space–weather gauntlet \citep{luger2015b,airapetian2017,noack2017a}.

%%%%%%%%%%%%%%%%%%%%%%%%%%%%%%%%%%%%%%%%%%%%%%%%%%%%%%%%%%%%%%%%%%%%

\section{Discussion}
\label{discussion}

%%%%%%%%%%%%%%%%%%%%%%%%%%%%%%%%%%%%%%%%%%%%%%%%%%%%%%%%%%%%%%%%%%%%

\subsection{Lessons for Interpreting Rocky Exoplanets}

Studying Mars provides numerous lessons that translate directly into
the interpretation of rocky exoplanet data. Regarding escape-driven
desiccation and abiotic O$_2$, long-term H escape from H$_2$O can
leave O$_2$/O$_3$-rich atmospheres without the need to invoke
biological processes. However, it is important to note that coupled H
and O escape may prevent significant O$_2$ accumulation in the
atmosphere \citep{mcelroy1972a,jakosky2018b}, and thus the degree to
which abiotic O$_2$ can build up depends on the balance between
production and loss channels. Mars further constrains rates and
controlling physics across dust seasons, informing false-positive
assessments around M-dwarfs \citep{heavens2018}. Mars provides
empirical tests of general circulation models (GCMs) and analytic
criteria for CO$_2$ freeze-out, and these results inform models of
tidally locked exoplanets in/near the HZ, where atmospheric mass and
heat transport determine survivability \citep{wordsworth2015a}, though
caution is warranted when extrapolating GCMs validated under
present-day Martian conditions to dynamically distinct regimes where
boundary conditions and processes may differ substantially. The
noble-gas and hydrogen isotope fractionation under known escape
regimes calibrate how we read future high-precision spectra of small
exoplanet atmospheres \citep{mahaffy2013}. Mars also informs the
potential for habitable windows in the absence of plate
tectonics. Sedimentary facies at Gale/Jezero demonstrate that
long-lived surface waters can exist transiently, even as a planet
trends toward aridity—guiding expectations for temporal habitability
on small exoplanets \citep{grotzinger2014b}.

%%%%%%%%%%%%%%%%%%%%%%%%%%%%%%%%%%%%%%%%%%%%%%%%%%%%%%%%%%%%%%%%%%%%

\subsection{Recommended Search Program for Mars Analogs}

As detailed in Section~\ref{det}, the detection of Mars analogs is a
challenging task as the required sensitivity lies at the edge or
beyond for the majority of exoplanet detection techniques. The most
promising near-term discovery pathways include transit+TTV
observations of compact systems around late M-dwarfs
\citep{gillon2017a,agol2021} and the Roman microlensing survey of
bulge fields \citep{penny2019}. The long-term goals are to
characterize the atmospheres of low-mass planets, empirically modeling
their evolutions from early degassing stages through to later more
tenuous states. Validation of retrieval techniques based on
disk-integrated spectra of Solar System planets have provided
invaluable insight into the requirements of achieving such goals
\citep{kane2021d,robinson2023}. Combining thermal-infrared
observations of the nearest Mars-analog exoplanet candidates
\citep{wolfe2024}, reflected-light imaging with HWO-class facilities,
and mid-infrared interferometry with concepts such as LIFE
\citep{quanz2022a} will ultimately quantify the occurrence rate and
atmospheric properties of true Mars analogs, although confirmation of
their exact mass, composition, and habitability will remain
observationally demanding.

%%%%%%%%%%%%%%%%%%%%%%%%%%%%%%%%%%%%%%%%%%%%%%%%%%%%%%%%%%%%%%%%%%%%

\subsection{Priority Topics for Mars Exploration}

In the most recent planetary science Decadal Survey, Origins, Worlds,
and Life\footnote{\url{https://doi.org/10.17226/26522}}, Mars features
heavily both as a worthy Solar System target in its own right and as a
basis for better understanding rocky exoplanets. Through a series of
Strategic Research (SR) activities within 12 Priority Science
Questions, key Solar System science topics can be addressed by
studying myriad aspects of the Mars system. For example, in Priority
Question 3, "Origin of Earth and Inner Solar System Bodies", SR 3.4
includes the objective to "Determine the formation time of Mars
through isotopic analyses of diverse Martian samples," something that
could be achieved through, for example, the return to Earth of samples
from the Red Planet. Similarly, in Priority Question 6, "Solid Body
Atmospheres, Exospheres, Magnetospheres, and Climate Evolution" there
are two SRs specific to understanding the atmospheric evolution of
Mars. One, SR 6.1, states "Constrain the earliest stages of
atmospheric evolution on Venus, Mars, and Titan by measuring noble gas
abundances and isotopic fractionation to sufficient precision to
quantify their minor isotopes." For the second, SR 6.2, the objective
is to "Determine how and why Mars’s climate has changed over orbital
timescales by performing radar and spectroscopic mapping of the polar
layered terrain and by making in situ measurements of their structure
and composition (thickness of layers, dust content, and isotope
ratios) and their local meteorology (including volatile and dust
fluxes)."

These are specific tasks that can be carried out at Mars or on Earth
with Mars samples, and are directly in service of better understanding
the modern state of the planet and how it got that way. In that same
Priority Question, however, SR 6.6 reads "Investigate the
microphysical parameters that influence the formation of clouds in
planetary atmospheres (primarily... Mars...) by determining the
distribution, nature, and abundance of clouds and the composition and
particle size of the droplets comprising them and cloud condensation
nuclei around which they form." This investigation would have direct
bearing on efforts to characterize rocky-exoplanet atmospheres in
terms of distinguishing potential cloud spectral signatures from that
of the bulk atmosphere.

Moreover, Origins, Worlds, and Life also includes an exoplanet-focused
chapter, written explicitly with the goal of using that understanding
of Mars to further our (remote) exploration of terrestrial
exoplanets. For instance, in the Priority Question 12, "Exoplanets",
SR 12.10 is to "Determine the properties of the atmospheres of
terrestrial planets (i.e., ... Mars) that would be observable on
exoplanets to build a foundation for atmospheric characterization of
analogue exoplanets through coordinating in situ/remote sensing
measurements and theoretical studies of wind velocities, radiative
balance, cloud dynamics, and atmospheric compositing as function of
orbital phase, local time, and solar conditions." SR 12.10 states
"Improve exoplanet habitability predictions for cold, low-mass planets
by determining the key factors that made Mars habitable 3–4 Ga, via a
combination of in situ geological and atmospheric analysis and sample
return, orbital observations, and climate modeling." And in SR 12.11,
the decadal survey advises "Study methods to discriminate past and
present false positive biosignatures on Solar System bodies (e.g.,
abiotic O$_2$ on... Mars) from true biosignatures to inform false
positive discrimination methods for exoplanets through in situ, remote
sensing, theoretical/modeling studies, analog field research, and
laboratory studies that characterize remotely observable properties of
these features."

Together, these activities if implemented would advance not only our
understanding of Mars' past and present states but would more firmly
place it as a foundation for how we identify and interpret Mars-analog
exoplanet features and properties.

%%%%%%%%%%%%%%%%%%%%%%%%%%%%%%%%%%%%%%%%%%%%%%%%%%%%%%%%%%%%%%%%%%%%

\section{Conclusions}
\label{conclusions}

Mars occupies an important position in comparative planetology, since
it is both a geologically rich world with a documented history of
surface habitability, and a representative example of how small rocky
planets can evolve toward atmospheric loss and climatic
decline. Through examination of its various properties, such as
formation history, orbital evolution, interior cooling, and
atmospheric escape, we have shown that Mars provides a physically
grounded framework for interpreting sub-Earth exoplanets. The
geological and atmospheric evolution of Mars underscores that
habitability is not a static property, but is a time-dependent outcome
governed by competing processes. Early volcanism and volatile release
likely supported a thicker atmosphere and surface liquid water, yet
declining interior heat flow, cessation of dynamo activity, and
sustained atmospheric escape gradually reduced surface pressure and
limited climate stability. These coupled processes can define a
pathway that may be common for Mars-mass planets. In this context,
Mars represents the edge of the habitable regime, being large enough
to host transiently clement conditions, but small enough that
atmospheric retention and replenishment and long-term climate
regulation are not guaranteed.

Our discussion of exoplanet demographics have shown that, while
terrestrial-size planets are abundant, confirmed Mars-mass planets
with well-constrained masses and radii remain relatively rare, largely
due to detection shortcomings. Transit and RV surveys favor larger or
more strongly irradiated planets, and the smallest rocky bodies
challenge current sensitivity limits. The coming Roman microlensing
surveys offer a particularly promising avenue for probing the true
frequency of Mars analogs. Direct imaging and thermal emission
studies, particularly with next-generation facilities, will ultimately
determine whether such planets commonly retain thin CO$_2$
atmospheres, undergo desiccation, or exhibit transient volatile
cycles.

Ultimately, the convergence of Solar System exploration and exoplanet
characterization provides a powerful strategy for understanding small
rocky worlds. Mars missions will continue to measure atmospheric
escape rates, volatile inventories, and climate feedbacks with a level
of detail unattainable for exoplanets, while exoplanet surveys
contextualize Mars within a broader statistical population. Together,
these complementary approaches identify key science priorities,
including understanding the mass threshold for sustained geologic
activity, quantifying atmospheric survival as a function of stellar
environment, assessing the stability of thin CO$_2$ atmospheres, and
refining observational techniques capable of detecting sub-Earth
planets. Within this framework, Mars provides a fundamental benchmark
for evaluating the diversity, evolution, and potential habitability of
rocky planets throughout the Galaxy.

%%%%%%%%%%%%%%%%%%%%%%%%%%%%%%%%%%%%%%%%%%%%%%%%%%%%%%%%%%%%%%%%%%%%

\section*{Acknowledgements}

The research described in this paper has benefited from useful
conversations with Brian Jackson. The authors also acknowledge the
valuable feedback provided by Bruce Jakosky and an additional
anonymous reviewer. This research has made use of the Habitable Zone
Gallery at hzgallery.org. The results reported herein benefited from
collaborations and/or information exchange within NASA's Nexus for
Exoplanet System Science (NExSS) research coordination network
sponsored by NASA's Science Mission Directorate.

%%%%%%%%%%%%%%%%%%%%%%%%%%%%%%%%%%%%%%%%%%%%%%%%%%%%%%%%%%%%%%%%%%%%

\software{Planetary Spectrum Generator \citep[PSG;][]{villanueva2018b}}

%%%%%%%%%%%%%%%%%%%%%%%%%%%%%%%%%%%%%%%%%%%%%%%%%%%%%%%%%%%%%%%%%%%%

%\bibliographystyle{aasjournal}
%\bibliography{/data/skane/latex/styles/references}

%%%%%%%%%%%%%%%%%%%%%%%%%%%%%%%%%%%%%%%%%%%%%%%%%%%%%%%%%%%%%%%%%%%%

\end{document}